# Tunneling Electroresistance Induced by Interfacial Phase Transitions in Ultrathin Oxide Heterostructures

*Lu Jiang[1,2,†], Woo Seok Choi[1,†], Hyoungjeen Jeen[1], Shuai Dong[1,2,3], Yunseok Kim[4], Myung-Geun Han[5], Yimei Zhu[5], Sergei V. Kalinin[4], Elbio Dagotto[1,2], Takeshi Egami[1,2,6], and Ho Nyung Lee[1*]*

[1]Materials Science and Technology Division, Oak Ridge National Laboratory, Oak Ridge, Tennessee 37831, United States

[2]Department of Physics and Astronomy, University of Tennessee, Knoxville, Tennessee 37996, United States

[3]Department of physics, Southeast University, Nanjing 211189, China.

[4]Center for Nanophase Materials Sciences, Oak Ridge National Laboratory, Oak Ridge, Tennessee 37831, United States

[5]Condensed Matter Physics and Materials Sciences Department, Brookhaven National Laboratory, Upton, New York 11973, United States

[6]Department of Materials Science and Engineering, University of Tennessee, Knoxville, Tennessee 37996, United States

ABSTRACT

The ferroelectric (FE) control of electronic transport is one of the emerging technologies in oxide heterostructures. Many previous studies in FE tunnel junctions (FTJs) exploited solely the differences in the electrostatic potential across the FTJs that are induced by changes in the FE polarization direction. Here, we show that in practice the junction current ratios between the two polarization states can be further enhanced by the electrostatic modification in the correlated electron oxide electrodes, and that FTJs with nanometer thin layers can effectively produce a considerably large electroresistance ratio at room temperature. To understand these surprising results, we employed an additional control parameter, which is related to the crossing of electronic and magnetic phase boundaries of the correlated electron oxide. The FE-induced phase modulation at the heterointerface ultimately results in an enhanced electroresistance effect. Our study highlights that the strong coupling between degrees of freedom across heterointerfaces could yield versatile and novel applications in oxide electronics.

In quantum electronics, an ultrathin insulating layer sandwiched by two metallic electrodes serves as a tunnel junction (TJ), where electrons pass through the barrier by the quantum tunneling process. Depending on the selection of electrode materials, a wide range of TJs can be designed and utilized for various purposes. For example, semiconductor tunnel diodes are used for amplifiers and frequency converters,[1] magnetic TJs for magnetic detectors in hard disks,[2] and superconducting TJs utilizing the Josephson effect for magnetometer applications.[3] Recently, an upsurge of interest has focused on electrically-induced large resistance changes by TJs with ferroelectric (FE) insulating barriers, displaying the so-called tunneling electroresistance (TER) effect (see Figure 1a).[4,5] The spontaneous polarization of the FE barrier is predicted to affect the tunneling behavior in FE tunnel junctions (FTJs) through the interface effect,[6] inverse piezoelectric response,[7,8] and modification of the band structure of the FE barrier,[9] resulting in a bi-stable state for non-volatile switching. Correlation between the FE polarization direction and the tunneling conductance due to the change of electrostatic potential across the FE layer has also been observed in FE oxide heterostructures.[4,10-14] This FE field induced modulation of the TER can be particularly interesting when the FE layer is combined with correlated electron oxides (CEOs) as the metallic electrodes. Up to now, however, the role of the electrode (the CEO layer in the heterostructure) has often been ignored or underestimated in explaining the TER effect. While an electronic phase modification in the electrode by a neighboring FE layer has been theoretically predicted in seminal work by Tsymbal and collaborators,[4,15] there have been only few experimental efforts to attest those predictions.[16,17] This is largely due to the absence of a direct FE control on both the electronic and magnetic states of the CEO layer, even though electric-field controlled modifications of the magnetism or magnetic-field dependent resistivity changes have been observed in oxide heterostructures.[18] Since the modification of the CEO state should naturally accompany drastic changes in the physical properties of the heterostructures, this approach encompasses great potential for novel device applications, which could effectively utilize CEO phase modifications by controlling the FE polarization (Figure 1a).

$La_{1-x}Sr_xMnO_3$ (LSMO) is a hole-doped CEO and an excellent model system to explore the above-described effects because its electronic and magnetic phases are highly susceptible to a small change in doping (Figure 1b).[19] Therefore, a strongly coupled phase modulation could significantly influence the TER magnitude. An ultrathin LSMO was chosen as the CEO layer to maximize the effect and to convincingly confirm the expected effects by investigating both its electronic and magnetic properties. We note that, while LSMO thin films have been frequently used as bottom electrodes in FTJs in some previous studies, their contribution to the junction transport has never been fully appreciated.[10,20,21] In this work, we systematically investigate the coupled phase modulation in the oxide heterostructures by varying the nominal composition $x$ in LSMO ($x$ = 0.20, 0.33 and 0.50). $PbZr_{0.2}Ti_{0.8}O_3$ (PZT) has been chosen as the FE layer owing to its large remnant polarization (typically, our highly polar PZT film has a remnant polarization, $P_r$ = ~80 $\mu C/cm^2$, without thickness dependent variations.).[22] It is also worth noting that a large polarization is indispensable to strengthening the coupling across the heterointerface without the dielectric breakdown typically found in non-FE-containing field effect devices.[23] Schematics of heterostructures and corresponding energy band diagrams are shown in Figure 1a to illustrate the main ideas. When the FE polarization is upward, the carriers (holes) in LSMO become accumulated due to the spontaneous electric field at the interface. Conversely, when the polarization is downward, a hole-depleted state is realized. In addition to the different shape of the PZT band due to the polarization reversal, the hole depleted/accumulated states yield a substantial change in the in-plane and out-of-plane transport properties in FTJs. Even more importantly, the hole depletion/accumulation could trigger the crossing of a phase boundary of LSMO (Figure 1b), providing us with another degree of freedom which could largely amplify or even reverse the TER effect.

The impact of the FE polarization on the electronic and magnetic state of the LSMO layer could be clearly observed from the in-plane transport and magnetic properties of our heterostructures, as summarized in Figure 2. Since it was impossible to switch the as-grown polarization on the macroscopic scale necessary for the in-plane measurements, we compared the results of various LSMO ultrathin films with (upward polarization) and without (no polarization) a PZT capping layer. As previously reported, the spatial extent of the ample FE polarization effect is limited to only a few nanometers.[24] Therefore, we focused on heterostructures with an ultrathin LSMO layer (5 nm). The polarization direction of the as-grown PZT was upward, so the LSMO film with PZT capping layer should have a hole accumulation state. This would shift the LSMO phase to the increased doping side, *i.e.* to the right in Figure 1b. The temperature dependent resistivity, $\rho(T)$, curves in Figure 2a show that $x = 0.20$ LSMO indeed exhibits a remarkable polarization-induced insulator-to-metal transition. In the case of the PZT/LSMO heterostructure, the drastically reduced $\rho(T)$ value, the increased ferromagnetic (FM) Curie temperature ($T_c$), and the increased magnetization coherently indicate the modification of electronic and magnetic states of LSMO from a FM insulator to a FM metal by the FE polarization. Note that the strong FE field effect control, yielding here, *e.g.* a 100,000 fold change in resistivity at 50 K, is solely related with the polarization of our PZT film.[24] Such a huge change in resistivity by FE polarization is amongst the largest ever reported.[16] On the other hand, Figure 2c shows results for $x = 0.50$ LSMO displaying an opposite trend to that of $x = 0.20$ LSMO: $\rho(T)$ now increases with the FE field induced doping. While it was rather difficult to determine $T_c$ from the suppressed temperature dependent magnetization $M(T)$ curves due to the antiferromagnetic (AFM) phase of the highly doped LSMO, the $T_c$ estimation from the peak in the $\rho(T)$ curves (denoted as triangles) shows a slight decrease in $T_c$ with the PZT layer on top of the LSMO layer. Note that the suppressed changes in $T_c$ compared to those of $x = 0.20$ LSMO might be due to the reduced slope value of the $x$-dependent $T_c$ curve, as shown in Figure 1b: the slope near $x = 0.50$ ($dT_c/dx \approx -800$) is only about 60% of the value near $x = 0.20$ ($dT_c/dx \approx 1300$). Based on the results for $x = 0.20$ and $0.50$ LSMO, it can be further expected that $x = 0.33$ LSMO should display less pronounced changes than the other compositions in the electronic and magnetic properties by the FE field induced doping. Indeed, as shown in Figure 2b, the changes in $\rho(T)$ due to the PZT capping are weaker when compared to the heterostructures with $x = 0.20$ and $0.50$ LSMO. The overall changes in the in-plane physical properties clearly indicate that the FE field doping is an efficient way to controlling the phase of the CEO layer,[16] which could substantially influence on the TER effect in the oxide heterostructure, as discussed later.

Figure 3 shows the polarization direction dependent tunnel currents, or TERs in a PZT/LSMO ($x = 0.20$) heterostructure. Figure 3a shows a piezoresponse force microscopy (PFM) phase image of the sample at room temperature, which was poled with an incrementally increasing $dc$ bias. The gradual change in the applied voltage used in switching the polarization direction shows that the coercive voltage is ~2 V. A conducting atomic force microscopy (CAFM) image for the corresponding area is shown in Figure 3b. The current map was collected at a tip bias voltage of −1.5 V. A distinct contrast has been observed in the current map, where the upward FE polarization of PZT leads to a significantly higher current. While a portion of the CAFM image includes noisy current spots due to instrumental limitations (see Supporting Information for more detail), the clear correlation between the PFM phase and CAFM images serves as a convincing proof that the FE polarization plays an important role in determining the tunneling current through the heterostructures.[10,11]

In addition, the overall tunneling behavior depends on the thickness of the ferroelectric layer.[25] At the thin limit, direct quantum tunneling can be expected, while Fowler-Nordheim tunneling governs the out-of-plane transport for samples with thicker PZT layers. More detailed discussions on the tunneling behavior can be found in Supporting Information. The current ratio in our heterostructure ($x$ = 0.20) between the upward and downward FE polarizations was obtained as a function of the PZT thickness, as shown in Figure 3c. It exhibited a peak around 7 nm of PZT. This behavior implies that there is an optimal thickness of the FE layer for the largest TER effect. When the thickness of the FE layer was too thin, the tunneling current was very large even for downward polarization, so the contrast seemed to be diminished. Also, as the PZT thickness approaches the critical thickness of ferroelectricity, the polarization decreases due to the enlarged effect of the depolarization field.[26,27] Since an increased polarization should result in a larger TER effect,[28] the decreased TER effect for thinner PZT might also be attributed to weakened ferroelectricity and increased leakage currents. On the other hand, when the PZT thickness is larger than ~10 nm, tunneling through the PZT layer would become extremely difficult, which again decreases the TER ratio. This reduction is caused by the exponential decrease of the tunneling probability as the thickness of the barrier layer increases. Note that the TER ratio for our PZT/LSMO heterostructure with the optimal PZT thickness (7 nm) is at least 30,000%. The actual value could be larger, but due to instrumental limitations, we calculated the ratio with the measurable minimum current level of 10 pA. While an accurate, apples-to-apples comparison of our TER with previously reported values is almost impossible due to the use of different combinations of FE and contact materials, testing method, sample thickness, bias voltage, etc., the TER ratio from our samples with correlated LSMO electrodes seems to be at least one order of magnitude larger than that from an almost identical PZT sample with our FTJs, but with highly conducting electrodes (Pt, SrRuO$_3$).[12] Consistent with our observation, we could also find a similar trend from related literature: The overall TER ratio from samples with CEO electrodes[10,13,14,21,29] is significantly larger than that from samples with non-CEO electrodes.[11,12,30]

As observed from the in-plane transport in Figure 2a, the modified phase of LSMO due to the FE polarization largely amplifies the electroresistance effect. Therefore, we believe that the LSMO CEO layer could play a more crucial role than the PZT FE layer for the electroresistance effect in the heterostructures. To examine this idea, we varied $x$ in our LSMO layer deposited prior to the PZT layer (10 nm in thickness). Figure 4 shows CAFM images for $x$ = 0.20, 0.33, and 0.50 in LSMO, taken at room temperature. A typical PFM phase image representative for all PZT/LSMO heterostructures confirmed the good ferroelectricity of PZT as shown in Figure 4a. It consistently displayed both positively and negatively poled domains with a clear contrast, regardless of $x$. On the other hand, the resultant TER effect observed from the CAFM image was strikingly different. For the $x$ = 0.20 heterostructure, as already shown in Figure 2, the upward (downward) polarization resulted in a significantly high (low) tunneling current. Surprisingly, for the $x$ = 0.50 heterostructure, this TER trend was completely reversed. As shown in Figure 4d, for $x$ = 0.50, the upward (downward) polarization resulted in a low (high) tunneling current, with a similarly distinct contrast in the current level as in the $x$ = 0.20 heterostructure. On the other hand, the $x$ = 0.33 heterostructure showed similar trends as the $x$ = 0.20 heterostructure, but the contrast between the current levels for the upward and downward polarizations was substantially suppressed.

In these regards, it is worthwhile mentioning again that the FE polarization in the PZT layer induces a hole-accumulated or -depleted state in LSMO near the interface. These changes in the hole carrier density near the interface could drastically alter the electronic and magnetic states of LSMO, especially when $x$ is near a phase boundary.[24] The distinctly different phase of LSMO would of course change the boundary condition at PZT/LSMO interface dramatically, influencing the tunneling current that we observe in the CAFM measurement. Note also that the $x$ = 0.20 and 0.50 LSMO heterostructures are in the vicinity of different phase boundaries. In fact, $x$ = 0.20 and 0.50 transit, respectively, from insulating and metallic phases to metallic and insulating phases by hole accumulation. Since the metallic (insulating) phase of LSMO would increase (decrease) the tunneling probability across the heterostructure, the opposite trend in the CAFM can be easily understood by taking into account the polarization-induced modulation of electronic states in LSMO. Furthermore, the $x$ = 0.33 LSMO is in the middle of the metallic phase which manifests that the FE field effect on the LSMO layer is weakened. Note that this suppressed contrast is in agreement with what most previous reports have observed with highly conducting electrodes.[11,12]

Some of the effects described here have been further verified by theoretical calculations. A microscopic model Hamiltonian was employed to simulate the depletion/accumulation of holes on the CEO side of the structure. More specifically, the Hamiltonian contains the standard two-orbital double-exchange (DE) term supplemented by an electrostatic potential originating from the surface charge of PZT that is expected to induce a redistribution of $e_g$ electrons in LSMO.[31] The FE polarization of PZT was modeled via a surface charge (±0.8 electrons/unit cell) attached to the LSMO interface, a reasonable and widely-accepted procedure.[31] Afterwards, the $e_g$ profile and screening potential in LSMO were calculated self-consistently by diagonalizing the DE Hamiltonian and solving the Poisson equation for the electrostatic field.[31] The localized spins in the DE model were assumed to be FM for $x = 0.20$ and 0.33, and A-type AFM (A-AFM) for $x = 0.50$, or paramagnetic (PM), *i.e.* disordered for all the values of $x$ studied here. Note that the results for the change in carrier populations were found to be quite similar leading us to conclude that the observed hole redistribution is dominated by electrostatics. As shown in Figure 5 and discussed previously, the screening effect on the LSMO layer is mostly restricted to just a few layers from the interface.[24] Although the screening length depends on the LSMO effective dielectric constant used, the qualitative tendencies are unambiguous: when the FE polarization points away from the LSMO layer, the interfacial $e_g$ density is prominently decreased (*i.e.* holes are accumulated). In contrast, the interfacial hole density is suppressed when the FE polarization points towards the LSMO layer. The FE modulation of the $e_g$ density near the interface is expected to modify the interfacial physical properties of LSMO significantly, in accord with well-established Mn-oxide theoretical phase diagrams and also with a recent prediction.[32] This interfacial phase transition will induce drastic changes in both the in-plane conductance and the out-of-plane tunnel current [15]. In fact, considering the $x = 0.50$ heterostructure as an example and using proper superexchange coupling, our calculation shows that the interfacial LSMO layers may become FM instead of the original A-AFM ground state order.[32] The possible interfacial FM to A-AFM phase transition in the $x = 0.20$ heterostructure will give rise to the opposite effect. Note that our calculations were performed for the ground state, namely at zero-temperature. Nevertheless, the results are overall in good agreement with the experimental observations. Moreover, since the amount of hole doping is similar for FM, A-AFM, and PM spin configurations, our results are useful to understand the large TER ratio of the FE/CEO junctions (see Supplementary Information for the PM result). The substantial modifications in the $e_g$ electronic density observed in our simplified model system indicate that the local phase transitions in LSMO near the interface can indeed be potentially dramatic, and this effect could be utilized for amplified electroresistance or other electronic devices that require large ON/OFF ratios.

Our combined experimental and theoretical observations robustly indicate that the modulated phase in LSMO is the key factor to control the TER, but the following additional details should also be considered for a deeper understanding of the system. First, the thickness of the CEO layer should be carefully controlled, in addition to the PZT thickness, as it significantly influences the current ratio. As the LSMO thickness approaches the ultrathin limit, LSMO tends to exhibit the behavior of a film with a smaller doping.[33,34] Since our LSMO ultrathin films are only 5 nm thick, they could behave as less-than-nominally doped films. (Note that the thickness dependence could vary with the doping level $x$ in LSMO) Although it would be rather difficult to quantify this thickness influence, the effect would simply shift the LSMO layer to a lower doping value in the vicinity of the phase boundaries of $x = 0.20$ and $0.50$ LSMO. Thus, this is not a serious problem. Second, the depletion width for the carrier depleted state for the downward PZT polarization should be considered as well. This would be especially important for future quantitative analyses, as the depletion width could change for different electronic phases of LSMO. Since the depletion width would directly affect the tunneling probability across the heterostructure, it is an important parameter to be considered. Third, due to the phase changes in the LSMO layer, the PZT layer can be inversely affected. For example, drastic modifications in the boundary condition could alter the depolarization field, which could directly affect the FE polarization. Finally, intrinsic phase separation tendencies in the manganite layers could also be an issue to consider in real materials.

In conclusion, we have demonstrated that TER in ultrathin FE/CEO (PZT/LSMO) heterostructures is determined by phase transitions in the interfacial state of LSMO induced by the FE polarization. This coupled phase modulation was confirmed using LSMO layers with compositions near the phase boundaries. The largest TER ratio was obtained for LSMO $x = 0.20$, reaching a ratio > 30,000%. This implies that the polarization induced phase transitions in the CEO layer play the most important role in determining the value of the TER. Our effort not only provides a comprehensive understanding of the electroresistance behavior in strongly coupled systems, but also contributes to the exploration of nanoscale highly sensitive non-volatile electronics, in which two different tunneling resistances define the logic states by the influence of the FE polarization.

**Methods.** All samples were *in-situ* grown by pulsed laser epitaxy on atomically smooth $TiO_2$-terminated $SrTiO_3$ substrates at 625 °C in 100 mTorr of oxygen. The growth at the high pressure oxygen partial pressure ensures the quality of our ultrathin heterostructures with chemically abrupt interfaces,[35] as confirmed by Z-contrast scanning transmission electron microscopy (Figure S6). A KrF excimer laser ($\lambda = 248$ nm) with a laser fluence of ~1 J/cm$^2$ was used for ablating sintered PZT and LSMO targets. Details on the growth condition can be found elsewhere.[22]

Nanoscale polarization and local conductance were mapped by PFM and CAFM, respectively. A clear FE response of hysteretic piezoresponse and switching behavior was observed in an as-grown FE film with an upward polarization direction. $\rho(T)$ curves were recorded by a 14 T physical property measurement system (PPMS, Quantum Design Inc.). Ohmic indium contacts were ultrasonically soldered to the samples' corners in Van der Pauw geometry and, then, gold wires were bonded to the contacts. $M(T)$ curves at 200 Oe were recorded using a 7 T superconducting quantum interference device (SQUID, Quantum Design Inc.) magnetometer. The structural quality of the heterostructures was investigated by x-ray diffraction and scanning transmission electron microscopy. For the later, we used the state-of-the-art JEOL ARM 200 CF, which was equipped with two aberration correctors (CEOS) and a cold field-emission gun, routinely achieving a spatial resolution of 0.8 Å. The range of collection angle of 68 to 280 mrad was used for high-angle annular dark-field imaging.

The model calculations were done using a 4×4×12 lattice with twisted boundary conditions (TBC) in plane, and open boundary conditions (OBC) perpendicular to the film. While the surface termination of LSMO may modify the size of the effective polarization at the PZT/LSMO interface as seen in a similar system,[36] a mixed surface termination (50%) of LSMO was used for the calculations to avoid the termination effect on the ferroelectric switching. The TBC, with a 6×6 $k$-mesh, can reduce finite size effects.[37] Several dielectric constants from 20 to 180 were used and all gave qualitatively similar results, although the screening length depends on the dielectric constant value. With regards to the $t_{2g}$ spins, an FM background for $x = 0.20$ and 0.33 and an A-AFM background for $x = 0.50$ were adopted to calculate the electronic distribution (Figure 5).[31] To account for the interfacial phase transitions, the system energies were compared between the original spin $t_{2g}$ backgrounds and those obtained by switching the background to the competing state, as done in Refs. 31 and 32. For example, in the $x = 0.50$ case with FE polarization pointing to the LSMO, four interfacial LSMO layers become FM when using $J_{AFM} = 0.1t_0$, where $J_{AFM}$ is the superexchange coefficient and $t_0$ is the DE energy unit (~0.5 eV).[19,31] Note that, in Figure 5, the changes in the hole density can be very large particularly for the first layer in our idealized calculation. However, our results should be considered only an upper bound on the density modifications that can be achieved by the influence of the FE component.

In particular, issues such as the lattice distortions in the vicinity of the interface have not been considered in our effort. For more details the reader should consult Ref. 31.

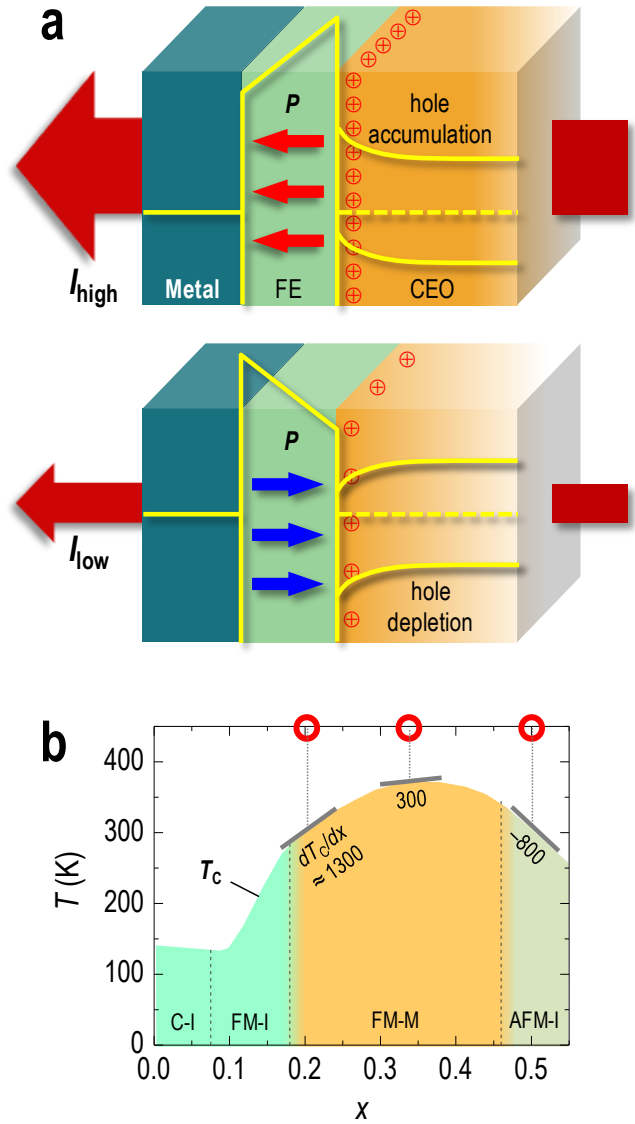

**Figure 1.** Interfacial phase modulation by FE. (a) Schematic representation of Metal/FE/CEO TJs. Energy band diagrams of FTJs at zero external bias are also drawn for two polarization directions. The carrier population is controlled by the direction of the polarization, which yields either hole accumulated (top) or depleted (bottom) state in the CEO layer. (b) Phase diagram of bulk LSMO varying $x$ [Ref. 19]. Near $x = 0.20$, $T_c$ changes the most with $x$ ($dT_c/dx \approx 1300$), near $x = 0.50$ the change is less drastic ($dT_c/dx \approx -800$), and near $x = 0.33$ it is the smallest ($dT_c/dx \approx 300$).

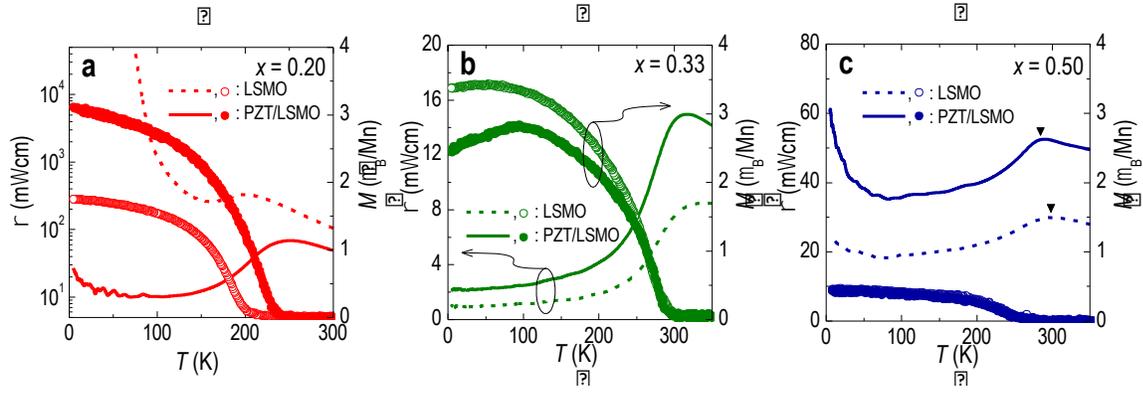

**Figure 2.** FE control of in-plane transport and magnetic properties. In-plane resistivity ($\rho$) (lines) and magnetization ($M$) (symbols) as a function of $T$. Results for ultrathin LSMO with compositions (a) $x = 0.20$, (b) 0.33, and (c) 0.50 are shown with and without PZT capping layers. Note that (a) is reproduced from Ref. 24

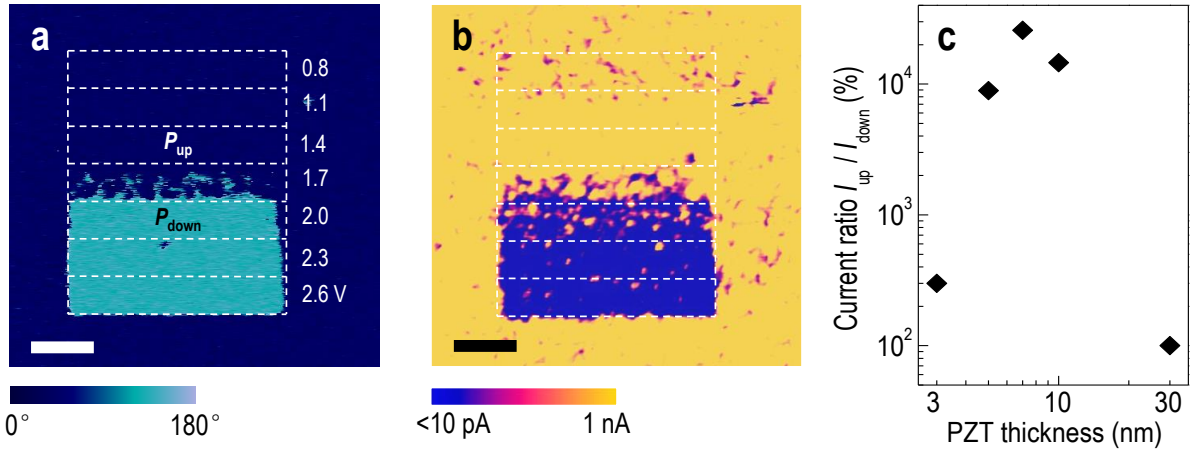

**Figure 3.** Polarization direction dependent tunneling. (a) Voltage-dependent PFM phase image of a PZT/LSMO ($x = 0.20$) heterostructure at room temperature. The thicknesses of PZT and LSMO are 10 and 5 nm, respectively. (b) CAFM map for the area corresponding to (a). Upward polarization results in higher current. In (a) and (b) the scale bars correspond to 300 nm. (c) Current ratio between $P_{up}$ and $P_{down}$ states as a function of the PZT thickness.

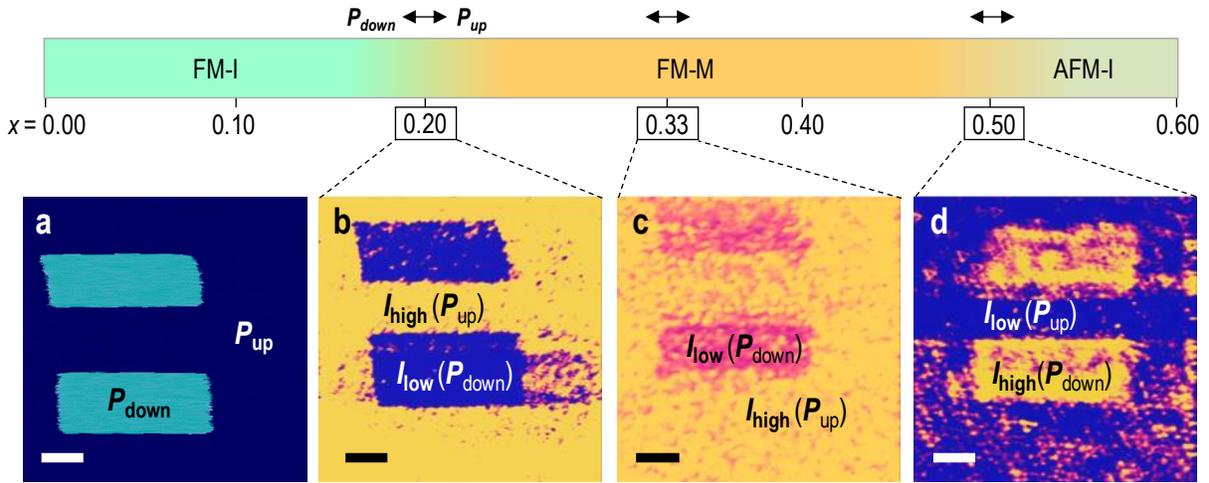

**Figure 4.** Tunneling current modulation at various phase boundaries. (a) PFM phase and (b to d) CAFM images of the PZT/LSMO heterostructure for compositions (b) $x = 0.20$, (c) 0.33, and (d) 0.50 at room temperature. The FE polarization induced phase transition in LSMO is shown via arrows. The scale bars on the images correspond to 300 nm. See Figure 3 for data scale bars.

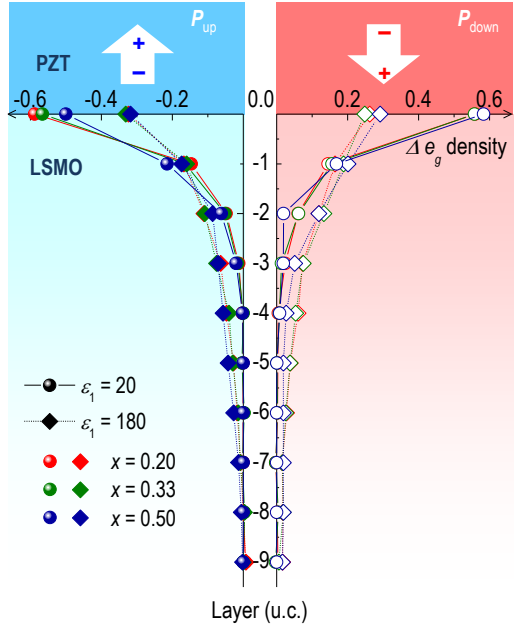

**Figure 5.** FE field control of interfacial charges. The theoretical depth profile showing the changes in the $e_g$ electronic density for upward ($Q = -0.8$) and downward ($Q = 0.8$) polarizations for compositions $x$ = 0.20, 0.33 and 0.50. The spin configurations in the DE model were selected as FM for $x$ = 0.20 and 0.33, and A-AFM for $x$ = 0.50. Two different dielectric constants ($\varepsilon_1 = 20$ and 180) have been used to compare the degree of FE control. These idealized calculations show that the electronic density modifications in the first and second layers can indeed be very large, compatible with the experimental results. Our theoretical results define an upper bound of what could be achieved experimentally, since effects such as lattice reconstructions at the interfaces have not been incorporated. In the Supplementary Information, results for PM spin configurations are presented as well, showing that the amount of hole doping does not change much with the spin configuration.

AUTHOR INFORMATION

**Corresponding Author**

*E-mail: hnlee@ornl.gov.

**Author Contributions**

†These authors contributed equally.

**Notes**

The authors declare no competing financial interest.

ACKNOWLEDGMENT

This work was supported by the U.S. Department of Energy (DOE), Basic Energy Sciences (BES), Materials Sciences and Engineering Division under contract numbers DE-AC05-00OR22725 (ORNL) and DE-AC02-98CH10886 (BNL). A portion of the PFM work was conducted at the Center for Nanophase Materials Sciences, which is sponsored at Oak Ridge National Laboratory by the Scientific User Facilities Division, DOE-BES. The STEM sample preparation was carried out at the Center for Functional Nanomaterials, Brookhaven National Laboratory. S.D. was in part supported by NSFC(11274060)

REFERENCES

(1) Esaki, L. *Phys. Rev.* **1958,** 109, (2), 603-604.

(2) Julliere, M. *Phys. Lett. A* **1975,** 54, (3), 225-226.

(3) Josephson, B. D. *Phys. Lett.* **1962,** 1, (7), 251-253.

(4) Tsymbal, E. Y.; Kohlstedt, H. *Science* **2006,** 313, (5784), 181-183.

(5) Zhuravlev, M. Y.; Sabirianov, R. F.; Jaswal, S. S.; Tsymbal, E. Y. *Phys. Rev. Lett.* **2005,** 94, (24), 246802.

(6) Velev, J. P.; Duan, C.-G.; Belashchenko, K. D.; Jaswal, S. S.; Tsymbal, E. Y. *Phys. Rev. Lett.* **2007,** 98, (13), 137201.

(7) Kohlstedt, H.; Pertsev, N. A.; Rodríguez Contreras, J.; Waser, R. *Phys. Rev. B* **2005,** 72, (12), 125341.

(8) Bilc, D. I.; Novaes, F. D.; Íñiguez, J.; Ordejón, P.; Ghosez, P. *ACS Nano* **2012,** 6, (2), 1473-1478.

(9) Hinsche, N. F.; Fechner, M.; Bose, P.; Ostanin, S.; Henk, J.; Mertig, I.; Zahn, P. *Phys. Rev. B* **2010,** 82, (21), 214110.

(10) Garcia, V.; Fusil, S.; Bouzehouane, K.; Enouz-Vedrenne, S.; Mathur, N. D.; Barthelemy, A.; Bibes, M. *Nature* **2009,** 460, (7251), 81-84.

(11) Gruverman, A.; Wu, D.; Lu, H.; Wang, Y.; Jang, H. W.; Folkman, C. M.; Zhuravlev, M. Y.; Felker, D.; Rzchowski, M.; Eom, C. B.; Tsymbal, E. Y. *Nano Lett.* **2009,** 9, (10), 3539-3543.

(12) Kohlstedt, H.; Petraru, A.; Szot, K.; Rudiger, A.; Meuffels, P.; Haselier, H.; Waser, R.; Nagarajan, V. *Appl. Phys. Lett.* **2008** 92, 062907.

(13) Maksymovych, P.; Jesse, S.; Yu, P.; Ramesh, R.; Baddorf, A. P.; Kalinin, S. V. *Science* **2009,** 324, (5933), 1421-1425.

(14) Chanthbouala, A.; Garcia, V.; Cherifi, R. O.; Bouzehouane, K.; Fusil, S.; Moya, X.; Xavier, S.; Yamada, H.; Deranlot, C.; Mathur, N. D.; Bibes, M.; Barthélémy, A.; Grollier, J. *Nat Mater* **2012,** 11, (10), 860-864.

(15) Burton, J. D.; Tsymbal, E. Y. *Phys. Rev. Lett.* **2011,** 106, (15), 157203.

(16) Vaz, C. A. F.; Hoffman, J.; Segal, Y.; Reiner, J. W.; Grober, R. D.; Zhang, Z.; Ahn, C. H.; Walker, F. J. *Phys. Rev. Lett.* **2010,** 104, (12), 127202.

(17) Yin, Y. W.; Burton, J. D.; Kim, Y. M.; Borisevich, A. Y.; Pennycook, S. J.; Yang, S. M.; Noh, T. W.; Gruverman, A.; Li, X. G.; Tsymbal, E. Y.; Li, Q. *Nat Mater* **2013,** 12, 397-402.

(18) Ramesh, R.; Spaldin, N. A. *Nat Mater* **2007,** 6, (1), 21-29.

(19) Dagotto, E.; Hotta, T.; Moreo, A. *Phys. Rep.* **2001,** 344, (1–3), 1-153.

(20) Hambe, M.; Petraru, A.; Pertsev, N. A.; Munroe, P.; Nagarajan, V.; Kohlstedt, H. *Adv. Funct. Mater.* **2010,** 20, (15), 2436-2441.

(21) Pantel, D.; Lu, H.; Goetze, S.; Werner, P.; Kim, D. J.; Gruverman, A.; Hesse, D.; Alexe, M. *Appl. Phys. Lett.* **2012,** 100, (23), 232902-4.

(22) Lee, H. N.; Nakhmanson, S. M.; Chisholm, M. F.; Christen, H. M.; Rabe, K. M.; Vanderbilt, D. *Phys. Rev. Lett.* **2007,** 98, (21), 217602.

(23) Stadler, H. L. *Phys. Rev. Lett.* **1965,** 14, (24), 979-981.

(24) Jiang, L.; Choi, W. S.; Jeen, H.; Egami, T.; Lee, H. N. *Appl. Phys. Lett.* **2012,** 101, (4), 042902-4.

(25) Pantel, D.; Alexe, M. *Phys. Rev. B* **2010,** 82, (13), 134105.

(26) Sai, N.; Kolpak, A. M.; Rappe, A. M. *Phys. Rev. B* **2005,** 72, (2), 020101.

(27) Junquera, J.; Ghosez, P. *Nature* **2003,** 422, (6931), 506-509.

(28) Zhuravlev, M. Y.; Wang, Y.; Maekawa, S.; Tsymbal, E. Y. *Appl. Phys. Lett.* **2009,** 95, (5), 052902-3.

(29) Pantel, D.; Goetze, S.; Hesse, D.; Alexe, M. *ACS Nano* **2011,** 5, (7), 6032-6038.

(30) Zenkevich, A.; Minnekaev, M.; Matveyev, Y.; Lebedinskii, Y.; Bulakh, K.; Chouprik, A.; Baturin, A.; Maksimova, K.; Thiess, S.; Drube, W.; *Appl. Phys. Lett.* **2013**, 102, 062907.

(31) Dong, S.; Zhang, X.; Yu, R.; Liu, J. M.; Dagotto, E. *Phys. Rev. B* **2011,** 84, (15), 155117.

(32) Burton, J. D.; Tsymbal, E. Y. *Phys. Rev. B* **2009,** 80, (17), 174406.

(33) Kim, B.; Kwon, D.; Song, J. H.; Hikita, Y.; Kim, B. G.; Hwang, H. Y. *Solid State Commun.* **2010,** 150, (13–14), 598-601.

(34) Huijben, M.; Martin, L. W.; Chu, Y. H.; Holcomb, M. B.; Yu, P.; Rijnders, G.; Blank, D. H. A.; Ramesh, R. *Phys. Rev. B* **2008,** 78, (9), 094413.

(35) Choi, W. S.; Rouleau, C. M.; Seo, S. S. A.; Luo, Z.; Zhou, H.; Fister, T. T.; Eastman, J. A.; Fuoss, P. H.; Fong, D. D.; Tischler, J. Z.; Eres, G.; Chisholm, M. F.; Lee, H. N. *Advanced Materials* **2012,** 24, (48), 6423-6428.

(36) Hikita, Y, Nishikawa, M.; Yajima, T.; Hwang, Harold, H. Y. *Phys. Rev. B* **2009**, 79 073101.

(37) Salafranca, J.; Alvarez, G.; Dagotto, E. *Phys. Rev. B* **2009,** 80, (15), 155133.

SUPPORING INFORMATION

**Tunneling through PZT/LSMO heterostructures**

The conduction mechanism in our heterostructures is tunneling, as shown in Figure S1. This figure shows I-V curves for A, $x = 0.20$ and B, $x = 0.50$ PZT (10 nm)/LSMO heterostructures. In general, when the electric field is larger than 1 MV/cm, Fowler-Nordheim (FN) tunneling can play a dominant role in the electronic conduction of tunnel junctions.[S1-3] As shown in Figure S1, the FN tunneling in our sample, regardless of the LSMO composition, is clearly evidenced from a linear relation in the $\ln(J/E^2)$ versus $1/E$ plot (see the insets of Figure S1), where $J$ is the current density and $E$ is the electric field. We note that the voltage used to obtain our CAFM maps (-1.5 V) is far above the onset fields ($E_{FN}$) for FN tunneling, which were 1.11 and 1.39 MV/cm for the $x = 0.20$ and 0.50 samples, respectively.

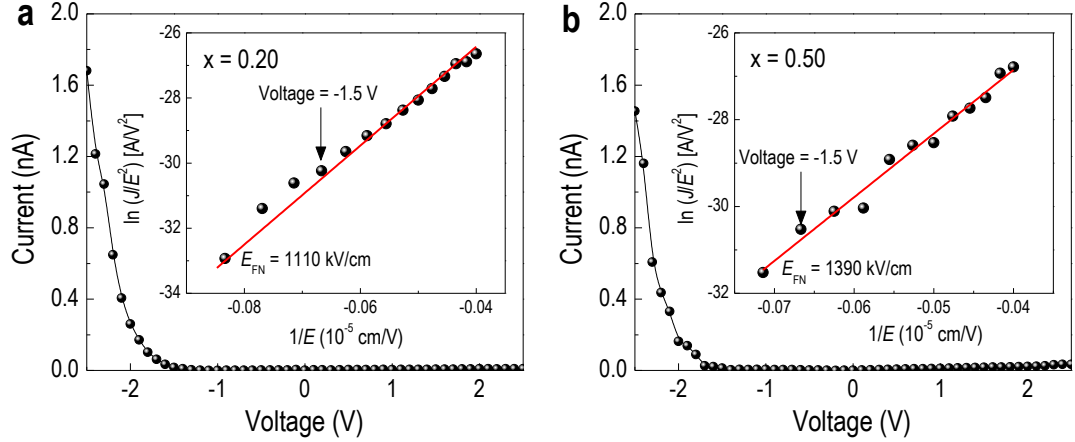

**Figure S1.** Voltage versus current curves. (a) $x = 0.20$ and (b) $x = 0.50$ PZT (10 nm)/LSMO heterostructures. The insets show a clear linear relation in the $\ln(J/E^2)$ versus $1/E$ plots, confirming the FN tunneling in the samples. The voltage used for CAFM measurements is well above the value $E_{FN}$, verifying that the conduction is mainly governed by tunneling.

In addition, it is worthwhile to discuss specific details related with the thickness of the heterostructures. Two different points need to be considered to address these issues properly: First, LSMO property dependent change (especially, in-plane resistivity), and, second, PZT thickness-dependent change, in tunneling probability.

Regarding the changes in tunneling as the properties of LSMO are modified, we first note that LSMO acts as one of the metallic electrodes. It would be difficult to quantitatively understand how the changes in the in-plane resistivity of LSMO would influence the out-of-plane junction transport. However, it is obvious that if the electric resistivity of the electrodes changes, the tunneling probability (or, overall junction conductance) of the heterostructure will also change,[S4] especially when the electrode is structurally isotropic such as in the case of LSMO. For example, if one of the electrodes changes simply from an insulator to a metal, it is apparent that the tunneling probability through the junction will increase substantially. We note that a similar reasoning can be applied to our samples, for example, the $x = 0.20$ heterostructure with a five-nm-thick-LSMO layer. Here, a huge change in the electrode resistance (and carrier density) is observed for opposite ferroelectric polarization directions. For thicker LSMOs, on the other hand, we found that the change in the polarization-dependent electric properties is much more suppressed, which leads to a smaller modification in the junction transport for the opposite polarization (see, Figure S2). Evidently, a PZT/LSMO heterostructure with a thicker LSMO (30 nm, Figure S2d) shows a much weaker TER contrast in the CAFM images compared to the one with a thinner LSMO (5 nm, Figure S2b).

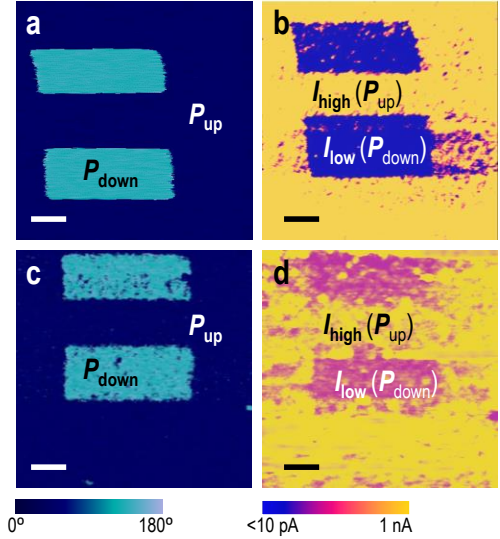

**Figure S2.** PFM and CAFM images of PZT/LSMO heterostructures with 5 nm and 30 nm thick LSMO. (a, c) PFM and (b, d) CAFM images with (a, b) 5 and (c, d) 30 nm thick LSMO. A smaller TER ratio is found using the heterostructure with a thicker LSMO layer compared to that with a thinner LSMO layer (weaker contrast in (d) compared to (b)), whose resistance change due to the ferroelectric field effect is much smaller. Note that (a) and (b) are the same images shown in the main text (Figure 4a and 4b).

Let us now focus on the PZT thickness-dependent changes in the tunneling probability.[S2] Note that PZT serves as a tunnel barrier here. The barrier thickness is one of the determining factors of the tunneling probability. From quantum mechanical tunneling, the tunneling probability decreases exponentially as the barrier thickness increases. As shown in Figure S3, both currents ($I_{low}$ and $I_{high}$) decrease rapidly with increasing the PZT thickness. For PZT thickness above 30 nm, the level of both currents approaches the measurement limit, and the difference between $I_{low}$ and $I_{high}$ becomes negligible. This indicates that the difference in the polarization-direction-dependent tunneling currents also decreases with increasing the PZT thickness. In addition, it is also worth mentioning that the TER ratio is also substantially reduced when the PZT layer is too thin ($\leq$ 3 nm). This phenomenon can be attributed to the ferroelectric size effect (i.e. disappearance of ferroelectricity below a certain thickness), as we have already mentioned in the manuscript, clearly validating our ferroelectric-polarization-driven TER control. Furthermore, this trend of TER ratio for ultrathin films is consistent with the previous report on $BaTiO_3$ with thickness < 3 nm.[S5]

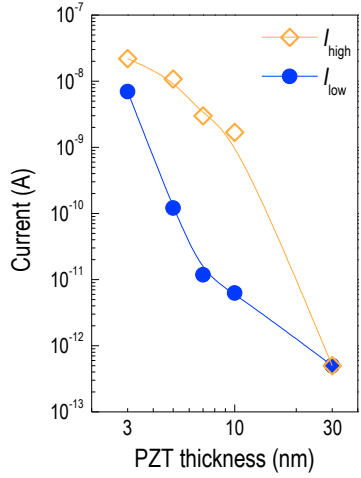

**Figure S3.** PZT thickness dependent tunneling currents for the $P_{up}$ ($I_{high}$) and $P_{down}$ ($I_{low}$) states in PZT/LSMO ($x = 0.20$) heterostructures.

**Influence of magnetic ordering on polarization-coupled interfacial carrier modulation**

To address the influence of magnetic ordering in LSMO on the charge carrier density modulation near the PZT/LSMO interface, we have additionally calculated the carrier concentration for other magnetic backgrounds in addition to the ferromagnetic order presented in the manuscript. Figure S4 shows the calculation data comparatively for both ferromagnetic and paramagnetic backgrounds, clearly indicating that the amount of doping near the interface does not strongly depend on the spin configurations. Note that the paramagnetic state of the LSMO layer was obtained by randomly selecting the orientation of the localized spin in the double-exchange model and, then, averaging over four configurations for each set of parameters. Moreover, the room temperature Fermi Dirac distribution was used.

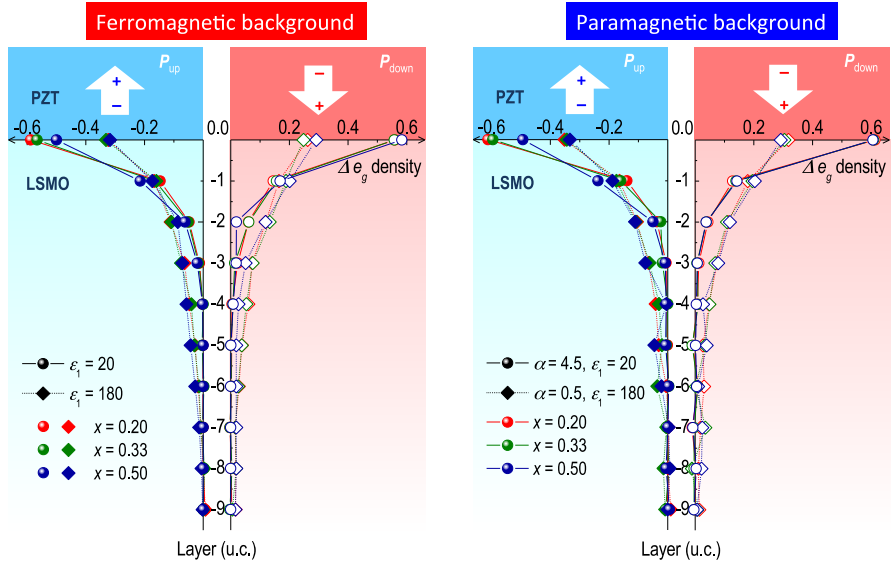

**Figure S4.** Carrier concentration profiles in the LSMO layer for ferromagnetic and paramagnetic backgrounds, near the vicinity of the PZT/LSMO interfaces.

It is worthy noting that there is a possibility of inducing a polar state in some of mananites due to orbital ordering,[S6,S7] even though our current approach is hard to understand if our manganite films also behave similarly.

**Mapping TER by CAFM**

Our CAFM data unambiguously revealed the important role of the coupling between the states (metallic or insulating) of the correlated electron oxide layer and the ferroelectric polarization to achieve a high TER. However, due to the instrumental limitations, a portion of the CAFM images includes noisy current spots. Here, we consider potential causes for the undesirable current spots in the CAFM mapping.

(1) *Electrochemical reactions*: Electrochemical reactions can occur during the application of a voltage since the PZT layer is on the top. However, it will be the same for all the LSMO compositions. Hence, we consider this contribution as an offset and, consequently, all the active variables reside at the PZT/LSMO interfaces. Furthermore, note that such electrochemical reactions should be enhanced when the applied voltage is increased, since it is an electric field dependent phenomenon. As shown in Figure 3, the current spots are not related to the applied voltage value, suggesting that the electrochemical reaction is not playing a role in our CAFM measurements. Also, the use of a metallic top electrode does not obviate the interfacial electrochemical reaction. Adversely, adding a top electrode will yield a completely different structural geometry since it could introduce an interfacial layer with extra parameters of top electrode materials.[S8,S9] Finally, we emphasize that any possible electrochemical reactions at our experimental condition could not reverse the sign of TER and should not change our main conclusion.

(2) *Switching current*: The current spots also appear in the as-grown region, where no ferroelectric switching is expected. In principle, if switching events occur during the CAFM measurement, a switching current could also appear in the as-grown region. However, even in such a case, the switching current is usually very small.

(3) *Locally different polarization in as-grown film*: This can be safely excluded based on our PFM images, which revealed a uniform response in our films.

(4) *Contact force*: While a large contact force could locally switch the ferroelectric polarization and can result in current spots,[S10] this cannot be the case here, as the ferroelectric switching is not the cause for the current spots, as discussed above in (2).

(5) *Current artefact or electric noise*: We believe this is the case for the current spots in our CAFM images. Unfortunately, our experimental setup cannot perfectly control the scanner, as also shown as distorted box images for both PFM and CAFM. For example, it is expected that locally unstable scanner control with a slightly damaged tip coating could cause the current spot problems for CAFM as similar to the case in CAFM measurements for rough surfaces.[S11]

(6) *Electromigration*: This phenomenon has been mainly observed from thick PZT films, usually beyond the tunneling limit. Therefore, we do not consider this effect in our junction transport.

**TER ratio reversal in capacitor geometry**

We have additionally conducted TER measurements using top-electroded capacitor geometry. We used a shadow mask (300 $\mu$m in diameter) to sputter-deposit Pt on top of PZT/LSMO heterostructures. As shown in Figure S5, the capacitor geometry revealed the same trend as in the tip-based measurements shown in Figure 4. For $x = 0.20$, the upward polarization state of PZT showed a higher current level than that for the downward polarization. On the other hand, for $x = 0.50$, this trend was opposite. While an accurate determination of the tunnelling current is hindered by the large leakage (or charging) current due to the huge RC time constant (intrinsic for such thin heterostructures), it clearly shows that the LSMO composition dependent TER behaviour is persistent even for different junction geometries.

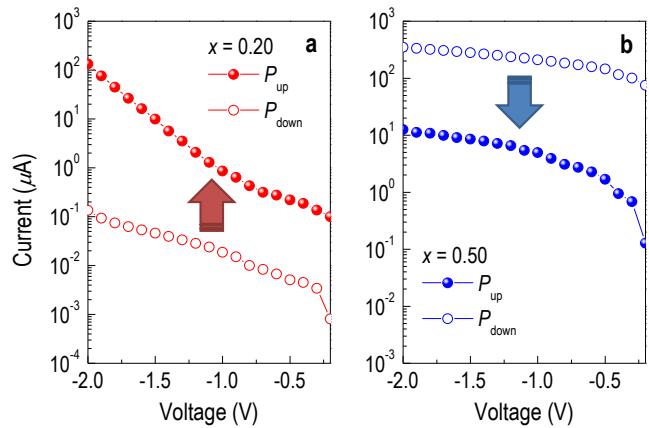

**Figure S5.** Voltage versus current curves PZT (10 nm)/LSMO heterostructures with Pt top electrodes. (a) $x = 0.20$ and (b) $x = 0.50$. The opposite TER trend is clearly seen from $x = 0.20$ and 0.50 consistent with the tip-based measurements shown in Figure 4.

**STEM investigation of the PZT/LSMO heterostructure**

We performed an atomic-resolution high-angle annular dark-field (HAADF) scanning transmission electron microscopy (STEM) investigation of a PZT/LSMO heterostructure to ensure the structural quality of our samples. Figure S6 shows cross sectional images of a PZT/LSMO heterostructure on a STO substrate. The low-resolution HAADF image in Figure S6a shows overall a good film quality and well defined interfaces. A closer look with high-resolution HAADF imaging (Figure S6b) and observed abrupt change across the interface in intensity profile for Pb and La columns (Figure S6c) clearly confirm that the PZT/LSMO interface is indeed atomically sharp.

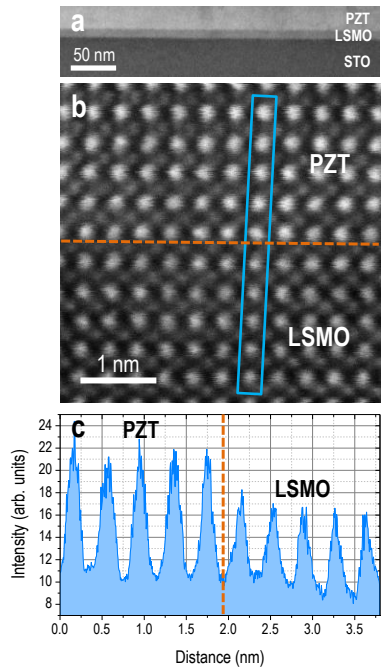

**Figure S6.** STEM HAADF images of a PZT/LSMO heterostructure on a STO substrate. (a to c) Low (a) and high (b) resolution images as well as the intensity profile (c) manifest a good quality of the PZT/LSMO interface.

**REFERENCES**
(S1) Fowler, R. H.; Nordheim, L. *Proc. R. Soc. Lond. A* **1928**, *119*, 173.
(S2) Pantel, D.; Alexe, M. *J. Appl. Phys.* **2005**, 98, 124104; *Phys. Rev. B* **2010**, 82, 134105.
(S3) Maksymovych, P.; *et al. Science* **2009** 324, 1421.
(S4) Will, T.; Escudero, R. *J. Appl. Phys.* **1981**, *52*, 1405.
(S5) Garcia, V. *et al.*, *Nature* **2009**, *460*, 81.
(S6) Burton, J. D.; Tsymbal, E. Y.; *Phys. Rev. Lett.* **2011** 106, 157203.
(S7) Ogawa, N. *et al.*, *Phys. Rev. Lett.* **2012** 108, 157603.
(S8) Bocher, L. *et al.*, *Nano Lett.* **2012**, *12*, 376.
(S9) Kim, D. J. *et al.*, *Nano Lett.* **2012**, *12*, 5697.
(S10) Lu H. *et al*., *Science* **2012**, *336*, 59.
(S11) Alexe M. *et al.*, *Adv. Mater.* **2008**, *20*, 4021.